\documentstyle[11pt,aaspp4]{article}
\tightenlines

\def\spose#1{\hbox to 0pt{#1\hss}}
\def\lta{\mathrel{\spose{\lower 3pt\hbox{$\mathchar"218$}}
     \raise 2.0pt\hbox{$\mathchar"13C$}}}
\def\gta{\mathrel{\spose{\lower 3pt\hbox{$\mathchar"218$}}
     \raise 2.0pt\hbox{$\mathchar"13E$}}}

\def\clock{\count0=\time \divide\count0 by 60
     \count1=\count0 \multiply\count1 by -60 \advance\count1 by \time
     \number\count0:\ifnum\count1<10{0\number\count1}\else\number\count1\fi}

\begin{document}


\title{The H$\alpha$ Luminosity Function and Global Star Formation
Rate From Redshifts of One to Two  } 

\bigskip

\author {Lin Yan\altaffilmark{1}, Patrick J. McCarthy\altaffilmark{1},}

\author{Wolfram Freudling\altaffilmark{2},
Harry I. Teplitz\altaffilmark{3,4}, Eliot M. Malumuth\altaffilmark{3,6},}

\author{Ray J. Weymann\altaffilmark{1}, Matthew A. Malkan\altaffilmark{5}}

\altaffiltext{1}{The Observatories of the Carnegie Institution of Washington, \\
                  813 Santa Barbara St., Pasadena, CA 91101}

\altaffiltext{2}{Space Telescope European Coordinating Facility,
                  Karl-Schwarzschild-Str. 2. D-85748 \\
                  Garching bei Munchen, Germany }

\altaffiltext{3}{NASA Goddard Space Flight Center,
                  Code 681, Greenbelt, MD 20771 }

\altaffiltext{4}{NOAO Research Associate}

\altaffiltext{5}{Astronomy Department, University of California,
                 Los Angeles, CA 90024-1562 }

\altaffiltext{6}{Raytheon STX Corp., Lanham, MD 20706}

\begin{abstract}
 
   We present a luminosity function for H$\alpha$ emission from
galaxies at redshifts between 0.7 and 1.9 based on slitless spectroscopy
with NICMOS on HST. The luminosity function is well fit by a Schechter
function over the range $6 \times 10^{41} <$ L$(H\alpha$) $ < 2 \times 10^{43}$
erg sec$^{-1}$ with L$^{*} = 7 \times 10^{42}$ erg sec$^{-1}$ 
and $\phi^* = 1.7
\times 10^{-3}$ Mpc$^{-3}$ for H$_0=50$~km s$^{-1}$ Mpc$^{-1}$ and q$_0=0.5$. 
We derive a volume averaged star formation rate
at $z = 1.3 \pm 0.5$ of 0.13 M$_{\odot}$ yr$^{-1}$ Mpc$^{-3}$ without correction
for extinction. The SFR that we derive at $\sim 6500$~\AA\
is a factor of $3$ higher than that deduced from 2800~\AA\ continua.
If this difference is due entirely to reddening, the extinction correction
at 2800\AA\ is quite significant.
The precise magnitude of the total extinction correction at rest-frame
UV wavelengths (e.g. 2800\AA\ and 1500\AA) is sensitive to
the relative spatial distribution of the stars, gas and dust, as well as on the
extinction law. In the extreme case of a homogeneous foreground dust screen and a MW or LMC 
extinction law, we derive a total extinction at 2800\AA\ 
of 2.1~magnitudes, or a factor of 7 correction to the UV luminosity density.
If we use the Calzetti reddening curve, which was derived for 
the model where stars, gas and dust are well mixed 
and nebular gas suffers more extinction than stars, 
our estimate of A$_{2800}$
is increased by more than one magnitude.

\end{abstract}


\section{Introduction}

The global star formation history of the universe holds clues to
understanding the formation and evolution of
galaxies.  The recent detections
of dust enshrouded galaxies at $z > 1$ at sub-millimeter
wavelengths (Smail, Ivison \&\ Blain 1997; Hughes et al. 1998; 
Barger et al. 1998; Lilly et
al. 1998) suggest that significant amounts of star
formation activity at high redshifts may be obscured. UV-selected
galaxies most likely represent only one segment of the population of active
star-forming systems at early epochs. The difficulty in estimating the
degree of extinction in the rest-frame UV introduces significant
uncertainties in the star formation history 
(Pettini et al. 1998;
Heckman et al. 1998; Blain et al. 1999; Malkan 1998).

Little is known about the properties of normal galaxies in
the region between $ 1 < z < 2$, where neither the 4000\AA\ break nor
the Ly continuum break are easily accessible.  The determinations of
the global star formation history based on either the 2500\AA\ and
1500\AA\ continuum luminosities, or on the abundance of damped 
Ly$\alpha$ systems (e.g. Pei \& Fall 1995; Madau et al 1996; Steidel et
al. 1998, Pei, Fall, \& Hauser 1999),
imply either a peak or a plateau in the $1 < z < 2$ range.
The near-IR offers one means of accessing both redshift indicators and
measures of star formation at these redshifts.

NICMOS offered a unique opportunity to perform slitless spectroscopy
in the near-IR. The extremely low background achieved on HST at
$\lambda < 1.9\mu$ allows for sensitive surveys for H$\alpha$ to $z =
1.9$.  We carried out a survey of random fields with the
slitless G141 grism ($\lambda_c = 1.5\mu$, $\Delta\lambda=0.8\mu$),
covering a total $\sim 64$ square arc-minutes (McCarthy et al. 1999;
hereafter paper I).  Our survey has equal or greater depth than 
current ground-based narrow-band imaging programs, and due to its large
wavelength coverage probes an order of magnitude more co-moving volume.
The details of the survey and the spectra of the
emission-line objects are given in paper I.

In this $Letter$ we present the H$\alpha$ luminosity function at $ 0.7 < z <
1.9$ derived from the H$\alpha$ emission-line galaxy sample
described in Paper I.  We also discuss the implications of our results
in the context of galaxy formation and evolution. Throughout this
paper we use H$_0=50$~km s$^{-1}$ Mpc$^{-1}$ and q$_0=0.5$.

\section{The H$\alpha$ Luminosity Function at $z \sim 0.7 - 1.9$}

\medskip
\subsection{The properties of the sample}

The bandpass of the G141 grism in camera 3 of NICMOS limits the range
of redshifts within which we can detect H$\alpha$ to $0.7 - 1.9$. 
The identification of the emission lines
as H$\alpha$ is based on H-band apparent continuum magnitudes, their
equivalent widths and the lack of other detected
lines within the G141 bandpass.  
In two of the three cases for which optical spectroscopy has been attempted with
LRIS on the Keck 10m telescope by Teplitz et al. (1999), the line identifications
have been confirmed by detection of their [OII]3727 emission.
These objects (J0055+8518a, $z=0.76$, J0622-0018a, $z=1.12$) are at the
low redshift end of our sample and therefore are the most accessible to
optical spectroscopy.
In the third case (J1237+6219A, $z=1.37$)
no emission or absorption features were detected in the spectrum.

\bigskip \subsection{Calculation of the H$\alpha$ luminosity function
at $z \sim 0.7 - 1.9$}

We used the  $\rm 1/V_{max}$ method (Schmidt 1968) to compute the
luminosity function. Each galaxy is assigned a
volume, $\rm V_{max}$, equal to the volume within which  it could lie and
be detected by our survey. We calculated the maximum co-moving
volume for each galaxy in each field with its appropriate line flux
limit.  The maximum volume is defined as

\begin{equation} V_{max} = \Omega \times \int_{zmin}^{zmax} \left({c
\over H_0}\right)^{3} {1 \over (1+z)^3} {\left(q_0z +
(q_0-1)(\sqrt{1+2q_0z} - 1) \right) \over q^4_0 \sqrt{1+2q_0z}} ~~
dz~, \end{equation}

where $zmin$ and $zmax$ are the minimum and maximum redshifts. $zmin = 0.67$,
the lower spectral cutoff of the NICMOS G141 grism.
The maximum redshift at which an object can be detected,
$zmax$, is ${\rm min}(1.9, z2)$, where 
$z2$ is computed from $D_L(zmax) = D_L(z)
\sqrt{f(H\alpha) \over f_{lim}}$, and $D_L$ is the luminosity
distance at redshift $z$. $f(\rm H\alpha)$ and $f_{lim}$ are the
H$\alpha$ line flux and the flux limit, respectively.  
$\Omega$ is the solid angle subtended by a single NICMOS camera 3 image.  
Only the central portion of each grism image
samples the full range of wavelengths from $1.1\mu - 1.9\mu$,
and the left and right portions of the image sample restricted
spectral ranges. The field of view subtended by the central section is
$\sim 31^{''}\times 49^{''}$; the two sides of the image 
subtend roughly $10^{''}\times 49^{''}$ each.
Thus the volume must be calculated separately for
each portion of the detector. The grism sensitivity is a function of
wavelength and this influences the effective volume probed,
particularly at the ends of the spectral coverage, where the
sensitivity (and hence the available volume) fall off steeply. 

The estimate of the source density in a luminosity bin of 
width $\Delta(\log L)$ centered on luminosity $\log L_i$ is simply the
sum of the inverse volumes ($\rm 1/V_{max}$) of all the sources with
luminosities in the bin. The value of the luminosity function in that bin is

\begin{equation}
\phi(\log L_i) = {1 \over \Delta (\log L)} \sum_{\mid \log L_j - \log L_i \mid < \Delta(\log L)}
{ 1 \over V_{max,j}}
\end{equation}

where index $i$ labels luminosity bins and index $j$ labels galaxies.
The variances are computed by summing the squares of the inverse
volumes; the error bars of each luminosity bin are the square roots of
the variances.

The apparent luminosity function must be corrected for incompleteness in the
original source catalog. 
The approach we adopted is similar to that used in 
Yan et al. (1998).  We chose
several well-detected emission-line galaxies spanning a range of W$_{\lambda}$
in the form of 2D spectra.
We dimmed these template spectra by
various factors and added them to the real NICMOS grism images at
random locations.  We then applied the same detection procedure as
in the original analyses (see Paper I) to recover the template
spectra.  The use of random positions in the simulation allows us to
include incompleteness corrections caused by crowding and spatially
dependent errors in the sky subtraction and shading corrections. We
added the dimmed template spectra to the 2D grism images with a wide
range of sensitivities to simulate the true distribution of
limiting fluxes in our survey fields.
The final averaged detection rate provides a 
measurement of the dependence of the detection probability on the line
flux limits, as well as equivalent width for W$_{rest} > 50$\AA.
The incompleteness correction is applied
to each bin independently.  

\begin{figure}
\plotone{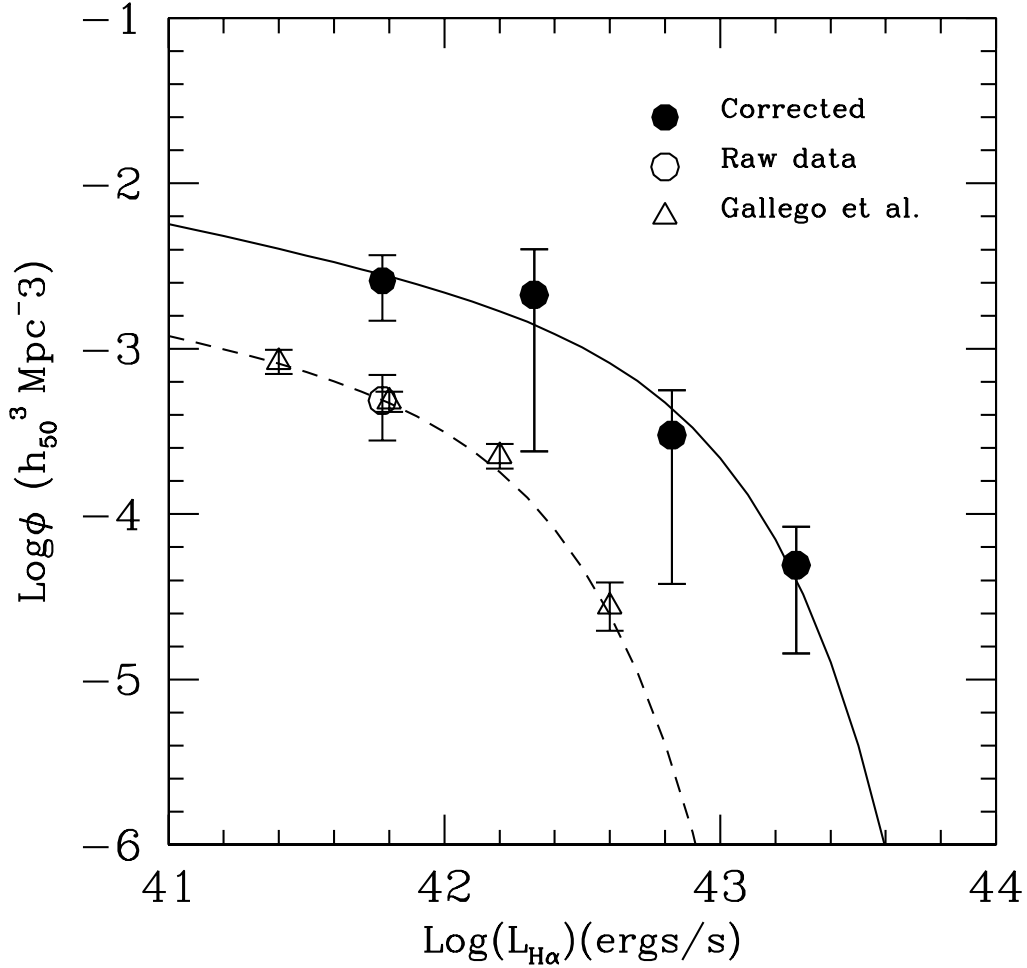}
\caption{H$\alpha$ luminosity function at $ 0.7 < z < 1.9$.
The open and filled circles are the data points from our measurements.
The open circles represent the raw data and the filled circles
are the points corrected for incompleteness. The incompleteness correction
is only significant at the faintest luminosity bin. The open triangles
show the local H$\alpha$ luminosity function by Gallego et al. (1995).
The solid and dashed lines are the best fits to the data
at $z \sim 1.3$ and $z \sim 0$ respectively.}
\end{figure}

In Figure 1 we show our derived luminosity function and the local H$\alpha$
LF as measured by Gallego et al. (1995).
Our H$\alpha$ luminosities have been corrected
for [NII] contamination using H$\alpha$/[NII]6583,6548 = 2.3
(Kennicutt 1992; Gallego et al. 1996).
The solid and dashed lines are the
best fits to a Schechter luminosity function at $z
\sim 1.3$ and $z \sim 0$, respectively. Gallego et al. (1995)
derived $\rm L^\star(H\alpha) = 1.4\times 10^{42}$~erg s$^{-1}$, $\alpha =
-1.35$ and $\phi = 6.3\times 10^{-4}$~Mpc$^{-3}$.  We
obtain $\rm L^\star(H\alpha) = 7\times 10^{42}$~erg s$^{-1}$ and $\phi =
1.7\times 10^{-3}$~Mpc$^{-3}$, assuming  a faint end slope, $\alpha$, equal to 
the local value of $-1.35$. Our sample is not large enough or deep enough to
allow an independent determination of $\alpha$. 

Figure 1 shows strong evolution in the H$\alpha$ luminosity density
from $z\sim 0 $ to $z \sim 1.3$.  This is no surprise given the evolution in
the ultraviolet luminosity density, but our result provides an independent
measure of evolution for H$\alpha$ emission alone.
The integrated H$\alpha$
luminosity density at $z \sim 1.3$ (our median $z$) is
$1.64\times 10^{40}$~h$_{50}$~erg s$^{-1}$ Mpc$^{-3}$, approximately 14 times
greater than the local value reported by Gallego et al. (1995). 
Gronwall (1998) has reported a preliminary measure of the local star formation
density derived from the KISS survey that is consistent with the Gallego et al. value.
If the faint end slope is as steep as
-1.6, as found for the  UV luminosity function at $z \sim 3$ (Steidel et al. 1999; 
Pascarelle, Lanzetta \&\ Fernandez-Soto 1998), the integrated H$\alpha$ luminosity density
at $z \sim 1.3$ would be roughly 50\% higher still.
Two of the emission-line galaxies in our sample are possible AGN candidates. If we
remove the top three objects with the highest H$\alpha$ fluxes from our luminosity function calculation,
the integrated H$\alpha$ luminosity density is reduced by $\sim$~40\%, primarily due to
the decrease of L$^\ast$(H$\alpha$). 

\section{Implications for the Evolution of Field Galaxies}

We converted the integrated H$\alpha$ luminosity density to a star
formation rate (SFR) using the relation from Kennicutt
(1999):  $\rm {SFR}(M_\odot yr^{-1}) = 7.9 \times 10^{-42} L(H\alpha)
(erg~s^{-1}) $.  This assumes Case B recombination at $T_e =
10^4$~K and a Salpeter IMF ($0.1 - 100~M_\odot$). This conversion
factor is about 10\% smaller than the value listed in Kennicutt (1983),
the difference reflecting updated evolutionary tracks.
While different choices of stellar tracks introduce modest uncertainties
in the conversion of UV and H$\alpha$ luminosities to star formation rates,
the choice of different IMFs lead to rather large differences. 
To make consistent comparisons between our results and those in the literature 
derived from 1500\AA\ and 2800\AA\ UV continuum luminosity
densities, we adopt the relation from Kennicutt (1999):
$\rm {SFR}(M_\odot yr^{-1}) = 1.4 \times 10^{-28} L(1500-2800\AA)
(erg~s^{-1} Hz^{-1}) $. This relation is appropriate for
the Salpeter IMF used to derive the H$\alpha$ conversion factor.

\begin{figure}
\plotone{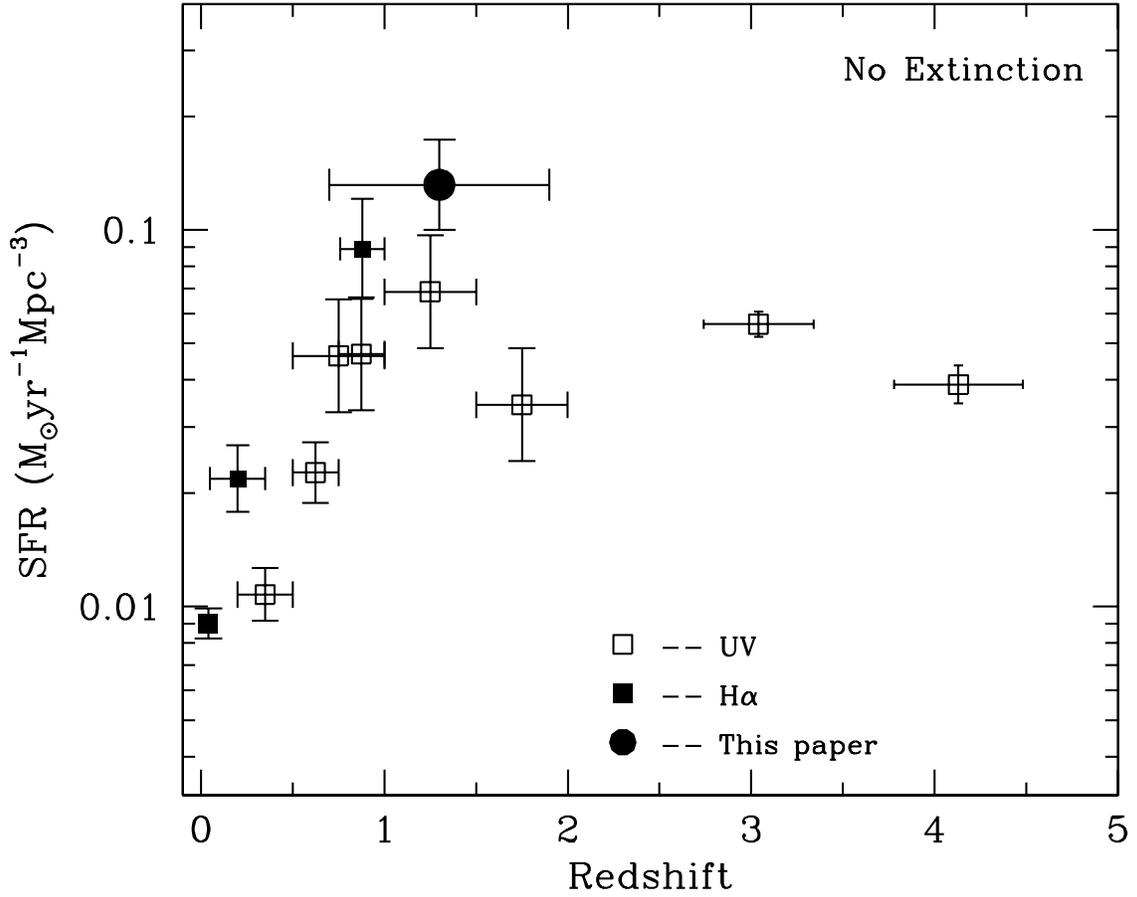}
\caption{The global volume-averaged star formation rate as a function of
redshift without any dust extinction correction. The open squares
represent measurements of the 2800\AA\ or 1500\AA\ continuum luminosity density by
Lilly et al. (1996), Connolly et al. (1996) and Steidel et al.  (1998),
whereas the filled squares are
the measurements using H$\alpha$~6563\AA\ by
Gallego et al. (1995), Tresse \&\ Maddox (1996) and Glazebrook et al. (1998).
Our result is shown in the filled circle.}
\end{figure}

In Figure 2, we plot uncorrected published measurements of the 
volume-averaged global star formation rate at various epochs. 
The open squares
represent measurements of the 2800\AA\ or 1500\AA\ continuum luminosity density by
Lilly et al. (1996), Connolly et al. (1996) and Steidel et al.  (1998);
the filled squares are the measurements using H$\alpha$~6563\AA\ by
Gallego et al. (1995), Tresse \&\ Maddox (1996) and Glazebrook et al. (1998).
Our result is shown as a filled circle. 
The star formation rates shown in Figure 2 are calculated from
the luminosity densities integrated over the entire luminosity 
functions, for both H$\alpha$ and the UV continuum.
Lilly et al. and Connolly et al. assumed a faint end slope of
$-1.3$ for the UV continuum luminosity functions at $\rm z \le 1$.
The 1500\AA\ continuum luminosity function at $z \sim 3 - 4$
measured from Lyman break galaxies by Steidel et al. (1998) has a 
faint end slope of $-1.6$.

Glazebrook et al. (1998) measured H$\alpha$ line fluxes
for 13 CFRS galaxies at $z \sim 1$ and
concluded that the star formation rate deduced from H$\alpha$
is significantly 
larger than that derived from the 2800\AA\ continuum luminosity density. Using 
the conversion factors which we have employed throughout this paper
($7.9\times 10^{-41}$ for H$\alpha$ and $1.4\times 10^{-28}$ for
the UV continuum), we estimate that without correcting
for extinction the star formation rate derived from their H$\alpha$ luminosity
density is a factor of 1.9 higher than that inferred from the 2800\AA\ continuum.
The factor of 3 difference in the star formation rate between H$\alpha$ and
the UV continuum noted in Glazebrook et al. (1998) includes an extinction
correction. As discussed below, the magnitude of the extinction correction 
is highly dependent on the relative spatial distributions
of stars, gas and dust.

Pettini et al. (1998) found the star formation rates
of Lyman break galaxies at $z \sim 3$ measured from H$\beta$ emission 
to be $0.7 - 7 \times$ higher than those estimated from the 1500\AA\
continuum. Hogg et al. (1998) and Hammer et al. (1997) derived the [OII]3727
luminosity function and detected strong evolution in the [OII] luminosity
density to $z = 1.3$. The conversion factor from the [OII]3727 luminosity
to a SFR, however, can differ by an order of magnitude, depending on the metallicity of the gas
(Gallagher, Hunter \&\ Bushhouse 1989; Kennicutt 1992).
In contrast, H$\alpha$ provides a more direct measure
of the star formation
by effectively reprocessing the integrated stellar 
luminosity of galaxies shortward of the Lyman limit.
Recent observations at sub-mm wavelengths indicate the star formation rate
at $2 < z < 4$ is larger than that inferred from the rest-frame UV luminosity density
(Hugh et al. 1998). The contribution of the sub-mm sources to the global star formation
history is uncertain at this time, as most of the sub-mm sources do not
have secure redshifts (Barger et al. 1999).

The clear trend for the longer wavelength determinations of
the star formation rate to exceed those based on UV continua is
one of the pieces of evidence for significant
extinction at intermediate and high redshifts. The amplitude of
the extinction correction is quite uncertain 
(e.g. Heckman et al.
1997; Meuers et al. 1997; Steidel et al. 1998). 
Our measurement spans $0.7 < z < 1.9$,
overlapping with the Connolly et al. photometric redshift sample and
allowing a direct comparison between the observed 
2800\AA\ luminosity density and that inferred from H$\alpha$.
The emission line galaxies selected by our survey have
a co-moving number density 
similar to that of the bright Lyman break galaxies
at $z \sim 3$, and a median H magnitude that corresponds to
approximately L$^\star$ (Paper I). While we are not comparing the
same individual galaxies, the rough correspondence in space density and
continuum absolute magnitudes between the UV- and H$\alpha$-selected samples
argues that they are drawn from similar or overlapping populations.
Our H$\alpha$-based star formation rate is three times larger than the
average of the three redshift bins measured by Connolly et al. (1997).

The star formation rates
derived from line or continuum luminosities depend strongly on the choice of 
IMF, evolutionary tracks, and stellar atmospheres that are input into a
specific spectral evolution model. The relevant issue for the present
discussion is the ratio of the star formation rates derived from
H$\alpha$ and the 2800\AA\ continuum. As shown by Glazebrook et al. (1998)
this ratio differs significantly for the Scalo and Salpeter IMFs and
is a function of metallicity. Our choice of the Salpeter IMF comes
close to minimizing the difference between the published UV- and our H$\alpha$-derived star
formation rates. 
The use of a Scalo IMF and solar metallicity would increase
the apparent discrepancy by a factor of $\sim 2$. The only model considered by
Glazebrook et al. that further reduces the H$\alpha$/2800\AA\ star formation
ratio is the Salpeter IMF with the Gunn \& Stryker (1983) spectral energy
distributions, and this model still leaves us with a factor of $\sim 2$
enhancement in apparent star formation activity measured at H$\alpha$.

If we attribute the entire difference to reddening, the total extinction corrections
at 2800\AA\ and H$\alpha$ are large and model-dependent.
The calculation is sensitive to the relative geometry 
of the stars, gas and dust, as well as the adopted reddening curve. 
In the extreme case of a homogeneous foreground screen
and a MW or LMC reddening curve, 
we derive A$_{2800} = 2.1$~magnitudes. 
In local starburst galaxies,
differential extinction between the
nebular gas, and stellar continuum, and scattering produce an effective reddening curve that is
significantly grayer than the MW or LMC curves
(Calzetti, Kinney \&\ Storchi-Bergmann 1994; 1996; Calzetti 1997).
The Calzetti reddening law (Calzetti 1997)
is appropriate for geometries in which the
stars, gas and dust are well mixed.
In this model, our estimate of the dust extinction
at 2800\AA\ is one to two magnitudes larger than in the simple screen case, and
is an uncomfortably large correction compared to results from other methods
(e.g., Heckman et al. 1998; Steidel et al. 1999).

The properties of the damped Ly$\alpha$ absorbers, 
diffuse backgrounds and galaxy counts at long wavelengths provide
independent constraints on the amount of obscured star formation at large redshifts
(e.g. Pei, Fall, \& Hauser 1999; Calzetti \&\ Heckman 1998; Blain et al. 1999).
Our measurement of the global star formation rate derived from
H$\alpha$ agrees well with the model predictions in Figure 7 of
Pei, Fall, \& Hauser. 

Despite our efforts to quantify the incompleteness of the
NICMOS grism sample, some biases remain. The low resolution
of the grisms prevent efficient detection of objects with 
line fluxes above our threshold but with rest-frame W$_{rest} < 50$\AA.
The Gallego et al. (1995) survey has a  W$_{\lambda}$ threshold of 10\AA.
Gallego et al. (1996) find that dwarf amorphous nuclear starbursts
have modest equivalent widths but contribute
little to the total luminosity density. Some of the compact starburst nuclei
will fall below our W$_{\lambda}$ threshold, and these objects
can have substantial luminosities. Their compact size mitigates against
this somewhat as our spectral resolution is best for point sources.
H$\alpha$ spectroscopy of galaxies from the CFRS sample at $z < 0.3$
by Tresse \&\ Maddox (1996) weakly suggests that  W$_{\lambda}$(H$\alpha$)
and luminosity are correlated. 

HST imaging and spectroscopic samples all suffer from a bias 
against low surface brightness objects.
The slitless nature of our survey exacerbates the problem
as the spectral resolution is a function of apparent source size.
The half-light radii
of the emission line galaxies in our sample range
from 0.2$^{''}$ to 0.7$^{''}$ and the distribution is comparable to
that seen in significantly deeper fields, such as the HDF-South and deep
NICMOS parallel fields (Yan et al. 1998; Storrie-Lombardi et al. 1999). 

The principal conclusions of this work are that the H$\alpha$ luminosity density at
$z = 1.3$ is an order of magnitude
larger than locally, that the global star formation rate derived from our H$\alpha$
measurements exceeds that from the rest-frame UV by a factor of 3 and the implied
extinction corrections are substantial. Although the characteristics of this particular 
data set do not lend themselves to precise comparison between global averaged
star formation rates inferred from UV continuum and line-emission, the systematically
larger rates inferred from H$\alpha$ at all redshifts point towards
significant extinction at rest-frame UV wavelengths.

\centerline{\bf 4. Acknowledgments}
\vskip 7pt
We thank the staff of the Space Telescope Science Institute for their
efforts in making this parallel program possible. In particular we
thank John Mackenty, Duccio Machetto, Peg Stanley, Doug van Orsow, and the staff of the PRESTO
division. We acknowledge useful discussions with M. Fall, J. Gallego,
D. Calzetti and R. Marzke. This research
was supported, in part, by grants from the Space Telescope Science
Institute, GO-7499.01-96A, AR-07972.01-96A and PO423101.  HIT acknowledges funding
by the Space Telescope Imaging Spectrograph Instrument Definition Team
through the National Optical Astronomy Observatories and by the NASA
Goddard Space Flight Center.

\end{document}